\documentclass{article}
\usepackage{amssymb}
\usepackage{amsmath}

\input{tcilatex}
\begin{document}

\title{Is the universe ill-posed?}
\author{John D. Barrow \\
DAMTP, Centre for Mathematical Sciences,\\
University of Cambridge, Cambridge CB3 0WA\\
United Kingdom}
\maketitle
\date{}

\begin{abstract}
We discuss an unusual consequence of the behaviour of general relativistic
cosmological models when their initial value problem is not well-posed
because of the lack of the local Lipschitz condition. A new type of 'zero
universe' arises with vanishing scale factor, and its first time derivative,
at all times. We discuss briefly some of the questions this raises about
creation out of nothing.
\end{abstract}

\section{Introduction}

The initial value problem for the Einstein equations of general relativity
proved to be a challenging mathematical problem due to the self-interacting
non-linearity of the system of partial differential equations coupled to the
space-time geometry, general coordinate covariance, and, like Maxwell's
equations, the presence of a set of constraint equations which govern
initial data. Key early contributions to a rigorous analysis were made in
classic work by Lichnerowicz \cite{lich}; for a comprehensive review of the
development, see Isenberg \cite{isen}. The aim of this ongoing work was to
establish that Einstein's equations, in vacuum and with well defined matter
sources are well posed: that good initial data determines the future
evolution of the space-time uniquely and completely. Great focus was laid on
these issues by the development of numerical relativity and by the Cosmic
Censorship hypothesis of Penrose \cite{pen}, as to whether naked
singularities can evolve from well-posed data; for recent results see \cite%
{mih}. New extensions of general relativity, like Horndeski's theories, have
also revealed that a well-posed initial value problem can be a significant
constraint on these theories \cite{JDBthor, reall}.

\ In cosmology, the uniqueness of the evolution from well-posed initial
value was used by Collins and Stewart \cite{CStew} to undermine the logic of
the chaotic cosmology proposal of Misner \cite{mis, mis1, mis2}, that
dissipative processes like neutrino viscosity could explain why almost all
anisotropic initial conditions could evolve to become as isotropic as the
newly discovered temperature isotropy of the cosmic microwave background
(then $\Delta T/T<10^{-3}$) after 10 billion years. Simply pick a universe
that meets present isotropy limits, and evolve it uniquely and continuously
backwards in time to find the required initial conditions. There will always
be a set. However, when looked at from a more physical perspective this does
not really work. The initial conditions required may be physically quite
unreasonable: for example, by requiring radiation or gravitational wave
energies to exceed to Planck density massively at the Planck time and so
quantum gravitational processes would intervene to equilibrate them \cite%
{JB2}, or they will have produced unacceptable levels of entropy in the
universe due to the dissipation of large anisotropies at very early times 
\cite{BM}.

These cosmological applications are of interest to us here because they make
essential use of the local Lipschitz condition as a constraint on the
cosmological evolution equations\footnote{%
Formally, a function $f:A\subset R^{n}\rightarrow R^{m}$ is locally
Lipschitz at $x_{0}\in A$ if there exist constants $ä >0$ and $M\in R$
such that
\par
$||x-x_{0}||<ä \implies ||f(x)-f(x_{0})||\leq M||x-x_{0}||$.
Informally, it requires trajectories that start close to stay close so
requires $f^{\prime }(x)$ to be bounded.}. \ 

\ In the next section of this short letter we look at a simple application
to Friedman universes that has an unusual consequence and then discuss the
issues raised by our results in the closing section.

\section{Indeterminate Friedman universes}

\bigskip Consider the Friedman metric,

\begin{equation}
ds^{2}=dt^{2}-a^{2}(t)\left\{ \frac{dr^{2}}{1-kr^{2}}+r^{2}[d\theta
^{2}+\sin ^{2}\theta d\phi ^{2}]\right\} ,  \label{ds}
\end{equation}

where $k$ is the curvature parameter, $t$ is comoving proper time, $r,\theta
,\phi $ are spherical polars, and $a(t)$ is the expansion scale factor
determined by the Einstein equations for (\ref{ds}) after specifying an
equation of state. Consider first a spatially flat universe with $k=0$. We
suppose that the scale factor has simple power-law time dependence, with

\begin{eqnarray}
a &=&t^{n},\dot{a}=nt^{n-1},\ddot{a}=n(n-1)t^{n-2}=n(n-1)a^{\frac{n-2}{n}},
\label{a} \\
\frac{d^{q}a}{dt^{q}} &=&n(n-1)(n-2)...(n-q+1)t^{n-q}\varpropto a^{\frac{n-q%
}{n}},
\end{eqnarray}

where overdot denotes $d/dt$. If we choose $n>1,$ then we can choose initial
conditions,

\begin{equation}
a(0)=\dot{a}(0)=0.  \label{init}
\end{equation}

We see that we have two very different solutions that both satisfy these
initial conditions but diverge to the future:

\begin{eqnarray}
a(t) &=&0,\text{ }\forall t  \label{b1} \\
a(t) &=&t^{n}  \label{b2}
\end{eqnarray}

This non-uniqueness phenomenon arises because the choice of $a=t^{n}$ with $%
n>1$ violates the local Lipschitz condition (as $\dot{a}$ is unbounded to
the future). This behaviour is not restricted to our power-law ansatz, (\ref%
{a}), as the choice $a=f(t),$ with $f(0)=f^{\prime }(0)=0$, for example with 
$f(t)=\sinh ^{n}(t)$, gives rise to the same non-unique behaviour if $%
f^{\prime }>0$. It has the same evolutionary behaviour as \ref{a} for $%
t\rightarrow 0$ but is de Sitter-like (inflationary) when $t\rightarrow
\infty $.

Our power-law example is not unusual as high power-law indices ($n>1$) are
needed for power-law inflation. The upside-down oscillator is an instructive
case and we can appreciate how inflation with a metastable equilibrium could
leave the scalar field there forever ($\dot{\varphi}=0$) or roll down the
potential. Both behaviours have the same initial conditions in the
metastable equilibrium but very different future evolution, like in eqs.(\ref%
{b1})-(\ref{b2}). \ 

A simple Newtonian and relativistic cosmological example is the universe
with scale factor $a=t^{n}$ that expands at constant power, $P\varpropto 
\ddot{a}\dot{a}\varpropto t^{2n-3}$: the power is constant when $n=3/2$ \cite%
{JB1}. Universes exerting constant force or acceleration, $F\varpropto \ddot{%
a}\varpropto t^{n-2}$, occur when $n=2$.

The simplest example we have is a flat Friedman universe with a perfect
fluid equation of state linking pressure, $p$, and density, $\rho $ by

\begin{equation}
p=(\gamma -1)\rho ,\text{ }\gamma \text{ constant.}  \label{b3}
\end{equation}%
The solution for the Friedman model is then $a(t)=t^{2/3\gamma }$, and so we
see that we get the behaviour leading to eq. (\ref{init}), corresponding to $%
n>1$, $\dot{a}>0$ as $t\rightarrow \infty $, when $\gamma <2/3.$ This Is by
no means an extreme or unusual state for matter ($\gamma =0$ is the vacuum, $%
p=-\rho $, state).

If we have $a=t^{n}$ in the flat Friedman model with no equation of state
linking $p$ and $\rho $, then the essential Einstein equations are ($8\pi
G=c=1$),$\ $

\begin{eqnarray}
3\left( \frac{\dot{a}^{2}}{a^{2}}\right) &=&\frac{3n^{2}}{t^{2}}=\rho
\label{fr1} \\
\frac{\ddot{a}}{a} &=&\frac{n(n-1)}{t^{2}}=-\frac{(\rho +3p)}{6}=-\frac{1}{2}%
\left[ \frac{n^{2}}{t^{2}}+p\right]  \label{fr2}
\end{eqnarray}%
and so

\begin{equation}
p=\frac{-n}{t^{2}}\left[ 3n-2\right] ,  \label{p}
\end{equation}

and 
\begin{equation}
\rho +p=\frac{2n}{t^{2}}  \label{prho}
\end{equation}

We note that $p=0$ for $n=2/3$ and $p=\rho /3$ for $n=1/2,$as expected. For
the solution with $a=0$ for all $t$ we have $\rho =p=n=0$, a type of
'nothing', or zero cosmology, where part of the metric still exists.

\section{Discussion}

We have shown that under certain easily realised (classical) conditions, the
Friedman equations has solutions like eq. (\ref{b1}) and (\ref{b2}) that
have identical initial conditions, $a(0)=\dot{a}(0)=0,$which lead to very
different future evolutions for $a(t)$. In particular, if we consider the
traditional description of 'creation out of nothing' in Friedman cosmology
then we assume the existence of spacetime with a metric like eq. (\ref{ds})
and regard $^{"}t=0^{"}$, where the density is infinity, as the beginning of
the universe. Expansion follows for $t>0$. No explanation for why the
universe comes into being, or starts expanding after $t=0$ is offered; nor
could it be, as this question is metaphysical and any answer to it lies
outside the encompass of relativistic Friedman cosmology. However, we have
displayed a new ingredient to this issue here. There are solutions of the
Friedman equation for isotropically expanding universes which share initial
conditions with solutions like (\ref{b1}), which never expand but just
remain forever in a unchanging state of zero expansion in which the space
portion of the metric vanishes while the time portion remains. Creation out
of nothing may create something that does not turn into a universe as we
understand it. What does this mean? Which evolutionary path does the
universe take from its $a=0=\dot{a}$ initial state and with what
probability? And are such zero-universe solutions stable?

\textit{Acknowledgement}. The author is supported by the Science and
Technology Facilities Council (STFC) of the United Kingdom.


\bigskip

\begin{thebibliography}{99}
\bibitem{lich} A. Lichnerowicz, \textit{L'integration des equations de la
gravitation relativiste et la probleme des n corps, }Journ. de Math. 23,
37-63, (1944).

\bibitem{isen} J. Isenberg, \textit{The Springer Handbook of Spacetime},
edited by A. Ashtekar and V. Petkov. (Springer-Verlag, New York)
arXiv:1304.1960

\bibitem{pen} R. Penrose, Riv. Nuovo Cim. 1, 252 (1969) and The Question of
Cosmic Censorship, in \textit{Black Holes and Relativistic Stars}, ed. R.
Wald, Univ. Chicago Press, Chicago, (1994) Chap. 5

\bibitem{mih} D. Christodoulou, Class. Quantum Grav. 16, A 23 (1999)

\bibitem{JDBthor} J.D. Barrow, M.Thorsrud and K. Yamamoto, JHEP 02, 146
(2013)

\bibitem{reall} R. Emparan and H. Reall, Living Reviews in Relativity 11,
article 6 (2008)

\bibitem{CStew} C.B. Collins and J.M. Stewart, Mon. Not. Roy. astron.
Soc.153 419 (1971)

\bibitem{mis} C.W. Misner, Phys. Rev. Lett. 19, 533 (1967)

\bibitem{mis1} C.W. Misner, Nature 214, 40 (1967)

\bibitem{mis2} C.W. Misner, Ap. J. 158, 431 (1968)

\bibitem{JB2} J.D. Barrow, Phys. Rev. D 51, 3113 (1995)

\bibitem{BM} J.D. Barrow and R. Matzner, Mon. Not. Roy. astron. Soc.181, 719
(1977)

\bibitem{JB1} J.D.Barrow, \textit{Mathletics, }Vintage, New York, (2013)
chap. 19.
\end{thebibliography}
\end{document}